\documentclass[12pt]{article}
\usepackage{amssymb,amsfonts,amsmath}
\usepackage{graphicx}
\usepackage{indentfirst}
\usepackage{fancybox}
\usepackage{color}

\parskip=\medskipamount

\def\f{\phi}

\def\z{\zeta}

\newcommand{\sect}[1]{\setcounter{equation}{0}\section{#1}}

\newcommand{\be}{\begin{equation}}
\newcommand{\ee}{\end{equation}}
\newcommand{\bea}{\begin{eqnarray}}
\newcommand{\eea}{\end{eqnarray}}

\def\double #1{#1{\hbox{\kern-2pt $#1$}}}

\begin{document}
\begin{flushright}
\end{flushright}

\begin{center}
{\Large \bf  Circumventing the No-Go Theorem} \\
\vspace{2mm}
{\Large \bf  in Noncommutative Gauge Field Theory}  \\
\end{center}

\begin{center}
\normalsize
{\large \bf  Masato Arai
\footnote{masato.arai@helsinki.fi},
Sami Saxell
\footnote{sami.saxell@helsinki.fi},
Anca Tureanu
\footnote{anca.tureanu@helsinki.fi}

and
Nobuhiro Uekusa
\footnote{nobuhiro.uekusa@helsinki.fi}
}
\end{center}
\vskip 1.2em
\begin{center}
{\it
High Energy Physics Division,
             Department of Physical Sciences,
             University of Helsinki \\
 and Helsinki Institute of Physics,
 P.O.Box 64, FIN-00014, Finland \\
}
\end{center}
\vskip 1.0cm
\begin{center}

{\large Abstract}
\vskip 0.7cm
\begin{minipage}[t]{14cm}
\baselineskip=19pt
\hskip4mm

Stringent restrictions
 for model building are imposed
 by a no-go theorem in noncommutative
 gauge field theory.
Circumventing this theorem is crucial for the construction of
 realistic models of particle interactions. To this end, the
 noncommutative construction of tensor representations of 
 gauge groups using half-infinite Wilson lines is
 extended to allow for gauge groups consisting of an arbitrary number
 of $U_*(N)$ factors.
This as well allows representations
 other than the ones permitted by the no-go theorem.

\end{minipage}
\end{center}

\newpage

\sect{Introduction}

The study of noncommutative (NC) gauge field theories has been
initiated from the early stages of the development of NC quantum
field theory, in connection with the observation that
 the noncommutativity of
space-time
 coordinates appears in string theory
 in the presence of an NS-NS $B$-field \cite{sw}. It was noted from
 the very beginning that the only allowed noncommutative gauge
 groups are the unitary groups. This is due to the fact that in NC
 field theory with Heisenberg-like commutation relation
\begin{equation}
\left[ \hat{x}^\mu, \hat{x}^\nu \right]=i\theta^{\mu\nu}, \label{NC}
\end{equation}
where $\hat{x}^\mu$ are the space-time coordinate operators,
 and $\theta^{\mu\nu}$ is an antisymmetric constant matrix,
the conventional procedure requires to
 replace the usual product between any fields with the
 Moyal star-product
\begin{equation}
(fg)(\hat{x})\longmapsto(f* g)(x) =
  \textrm{exp}\left[\frac{i}{2}\theta^{\mu\nu}
    {\partial\over \partial x^\mu}{\partial\over \partial y^\nu}\right]
f(x)g(y)
 \bigg| _{x=y}. \label{star}
\end{equation}
Due to the
Moyal star-product by which the gauge transformation are
 multiplied, a number of constraints arise, the first being that the
 closure condition is satisfied only by unitary groups $U_*(N)$ (see
 next section for details on their construction), while any
 other groups such as special unitary, orthogonal or symplectic gauge
 groups do not close. 
Two more consequences of the
 noncommutativity of the Moyal star-product deserve to be mentioned in this
 connection: i) unlike the commutative case,
 the unitary group $U_*(1)$
 is non-Abelian; ii) the group $U_*(N)$ is simple and not semi-simple
 as in the commutative case, however $U_*(1)$ is still a subgroup
 of $U_*(N)$, although the quotient $U_*(N)/U_*(1)$
 does not exist.

For the allowed $U_*(N)$ gauge groups, further restrictions appear
 \cite{Gracia-Bondia:2000pz, terashima, nogo}, gathered into a no-go
 theorem in Ref. \cite{nogo}. The theorem states that: 1) the local
 $u_*(n)$
 {\it algebra} only admits the irreducible $n\times n$ matrix
 representation. Hence the gauge fields are in $n\times n$ matrix
 form, while the matter fields can only be in fundamental, adjoint or
 singlet states; 2) for any gauge group consisting of several
 simple-group factors, the matter fields can transform nontrivially
 under {\it at most two} NC group factors. 
In other words, the matter
 fields cannot carry more than two NC gauge group charges.

Especially, the last restriction for charges is problematic for
 model building.
The construction of an NC version of the standard model (SM) \cite{CPST}
 disclosed the above problems.
Since only the unitary group is allowed as a gauge group, the natural minimal
 extension
 of the SM gauge group is
 $U_*(3)\times U_*(2)\times U_*(1)$.
With this choice, unlike in the commutative case, the quarks cannot have
 three gauge charges.
Since matter fields can only carry at most
 two gauge groups, the quarks can not be charged under the $U_*(1)$ group
 if they are charged under $U_*(3)$ and $U_*(2)$.
\footnote{It was shown that, provided that trace-$U_*(1)$ subgroups 
 of $U_*(3)\times U_*(2)\times U_*(1)$ are properly broken,
 the gauge boson of the residual $U_*(1)$ symmetry
 (corresponding to hypercharge $U(1)$ group in the commutative limit)
 couples to all the matter fields
 (placed into representations of $U_*(3)\times U_*(2)\times U_*(1)$
 strictly according to the no-go theorem) through the proper hypercharges
 (for further discussion, see \cite{CKT,AST}).}
If one wants to construct a model, in general,
 possessing gauge groups $\Pi_i U_*(n_i)$ and matter charged under more
 than two gauge groups, one would similarly encounter the restriction of the no-go
 theorem.
Therefore, it is crucial to circumvent the restrictions of the no-go
 theorem in a consistent way.

The progress in the formulation of NC gauge field theory 
 was not put off by the no-go theorem.
At different stages, steps were taken towards
 evading the requirements of this no-go theorem in NC gauge theory.
A key ingredient of the scheme is to
 satisfy the closure condition of direct product group by introducing
 an NC version of a Wilson line. NC Wilson lines were firstly
 introduced in the construction of NC gauge invariant operators
 \cite{Gross:2000ba}.
Then, an important step was to use
 a half-infinite NC Wilson line in order to
 construct tensorial representations
 of any rank for $U_*(N)$ in \cite{Chu:2001if,Chu:2002},
 where an NC extension of the supersymmetric $U_*(5)$
 Grand Unified Theory was proposed.

In this paper, we extend the use of the Wilson line in the NC
 gauge field theory to
 construct the action integral formed out of fields carrying any
 number of charges.\footnote{This
 construction was hinted at
 in the formulation of the NC minimal supersymmetric SM \cite{AST}.}
We first give a brief review of the no-go theorem \cite{nogo}.
Then we explain how tensorial representation for the $U_*(N)$
 gauge group is constructed by using the Wilson line, whose formulation
 is essential for our construction.
We also present symmetric and anti-symmetric representations
 by employing the formulation.
After that, we construct
 representation for direct product of an arbitrary number of gauge groups.
Finally we give a comment on a mechanism to break a trace $U_*(1)$
 part of a $U_*(N)$ gauge group proposed in \cite{CKT,AST}.

\sect{Group representations with half-infinite Wilson lines}
\subsection{ The no-go theorem in noncommutative gauge theory}

For the self-consistency,
 we give a brief account of the
 no-go theorem \cite{nogo} in NC
 gauge field theory.
Let us consider the
 NC version of gauge transformation
for a gauge field:
\begin{eqnarray}
 A_\mu \rightarrow U*(A_\mu - i\partial_\mu)*U^{-1}\,. \label{NC-gauge}
\end{eqnarray}
Here $U=e^{-i\lambda}_*$ is a gauge group element with insertion of
 the Moyal star-products between the matrix valued functions,
 and $\lambda=\lambda^aT^a$, where $T^a$ is the
 matrix for a representation of the gauge group.
The infinitesimal gauge transformation is
\begin{eqnarray}
 \delta A_\mu=\partial_\mu \lambda 
 -{i \over
  2}[T^a,T^b](\lambda^a*A_\mu^b+A_\mu^b*\lambda^a)
-{i \over 2}\{T^a,T^b\}(\lambda^a*A_\mu^b-A_\mu^b*\lambda^a)\,.
\end{eqnarray}
As one can see, this gauge transformation is not closed, i.e.,
 $\delta A_\mu$ is not Lie algebra-valued, unless
 $\{T^a,T^b\}$ is a linear combination of $T^c$.
For example, it is obvious that
 special unitary group does not satisfy the condition.
The only allowed gauge group is an NC version of
 unitary group, $U_*(N)$, for which the above requirement is automatically true.

In addition to these restrictions,
 the representations of the $u_*(n)$ Lie algebra
 are restricted to $n\times n$ hermitian matrices.
Hence the gauge fields are in $n\times n$ matrix form, while
 the matter fields can only be in fundamental ($F$),
 anti-fundamental ($\bar{F}$), adjoint ($F\times \bar{F}$)
 and bi-fundamental ($F\times \bar{F}^\prime$).
Furthermore, matter fields can
 only transform non-trivially under at most two simple subgroups of any
 gauge group consisting of a product of simple groups.
In other words, the matter fields cannot carry more than two NC gauge
 group charges.
For $U_*(1)$ this restriction means
 that the charges of the matter fields
 are quantized to just 0, $+1$ or $-1$ \cite{hayakawa}.

\subsection{Tensor representations of $U_*(N)$}\label{tensor-section}
An obvious requirement for the NC gauge group representations
is to satisfy
the closure property of the gauge group.
For the fundamental representation of the scalar matter field
denoted as the column
 vector $\phi^i$,
 the gauge transformation is defined by
\begin{eqnarray}
 \phi^i\rightarrow (\phi^U)^i=U^i_{~j}*\phi^j\,.
\end{eqnarray}
This satisfies gauge group multiplication law
\begin{eqnarray}
 (\f^U)^V=\f^{V*U}\, , \label{times}
\end{eqnarray}
where $V$ is another gauge group element.
One can also check that this property is satisfied for the anti-fundamental,
 adjoint and bi-fundamental representations of matter fields.
However, representations other than them such as higher rank
 tensorial representations are not allowed.
For instance, let us consider
 a rank-2 representation of the single gauge group $U_*(n)$, $\f^{ij}(x)$.
In this case, one can readily see that the
 NC gauge transformation for this field
\begin{eqnarray}
 \f^{ij}\rightarrow
 U^i_{~i^\prime}*U^j_{~j^\prime}*\f^{i^\prime j^\prime}\,, \label{gauge-trans}
\end{eqnarray}
 does not satisfy the group multiplication law (\ref{times}).

The construction of the tensorial
 representation was proposed in Ref. \cite{Chu:2002}. Since the basic
 ingredients of this construction are at the core of
the extension to a direct product of groups,
 we shall briefly review them here.

The idea is to modify
 the gauge transformation (\ref{gauge-trans}) 
 in a non-trivial gauge-field-dependent way so that the group
 multiplication law holds (\ref{times}).
Here we introduce the NC version of a half-infinite Wilson line,
\begin{eqnarray}
 W_C(x)&=&P_*\exp\left(ig\int_0^1
 d\sigma {d\z^\mu(\sigma) \over d\sigma}A_\mu(x+\z(\sigma))\right)\,, \label{wilson}
\end{eqnarray}
where the integration is along the contour $C$ from $\infty$ to $x$,
\begin{eqnarray}
 C=\left\{\z(\sigma)\,,0\le \sigma \le 1\,|\,\z(0)=\infty\,,\z(1)=0\right\}\,,
 \label{path}
\end{eqnarray}
and the path ordering
 involves the Moyal star-product between any functions.
Under the
 NC gauge transformation (\ref{NC-gauge}),
the Wilson line transforms as
\begin{eqnarray}
 W_C(x)\rightarrow U(x_1)*W_C(x)*U^{-1}(x_2)\,,
\end{eqnarray}
where $x_1$ and $x_2$ are endpoints of the contour.
Without loss of generality,
 we can restrict
 spatial components of $x_1$ to be at infinity, which we simply denote
 as $x_1\to \infty$.
Furthermore we restrict the allowed gauge transformation $U(x)$ to those
 which approach a constant $U_\infty$ as $x_1\rightarrow \infty$:
\begin{eqnarray}
 W_C(x)\rightarrow U_\infty W_C(x)*U^{-1}(x)\,. \label{W-trans1}
\end{eqnarray}
Note that the gauge transformation (\ref{NC-gauge})
 with boundary condition $A_\mu(x)\rightarrow 0$ as $x\rightarrow\infty$
 means $U_\infty= \textrm{constant}$.
We choose this constant as $U_\infty= 1$
\begin{eqnarray}
 W_C(x)\rightarrow  W_C(x)*U^{-1}(x)\, \label{W-trans}
\end{eqnarray}
by ignoring the global transformation at infinity which can be
 attributed to a normalization of the fields.

By using the NC Wilson line, let us find the modified
 gauge transformation law.
To do this, it is convenient to define the quantity
\begin{eqnarray}
 \Phi^{ij}=W_{C_1}{}^i_{~i^\prime}*W_{C_2}{}^j_{~j^\prime}*\phi^{i^\prime
  j^\prime}\,,\label{tensor2-inv}
\end{eqnarray}
where the subscripts $C_1$ and $C_2$ denote two contours
 which have the same endpoints (\ref{path}).
By requiring this quantity to be gauge invariant,
 one obtains the gauge transformation of $\f^{ij}$ as
\begin{eqnarray}
 \f^{ij}\rightarrow (\f^U)^{ij}
 =(U*W_{C_2}^{-1})^j_{~k}*U^i_{~l}*W_{C_2}{}^k_{~m}*\f^{lm}\,.\label{tensor2}
\end{eqnarray}
This gauge transformation satisfies the closure condition (\ref{times})
 \cite{Chu:2002}, so that it is a suitable gauge transformation.
In the $\theta^{\mu\nu}\to 0$ limit, the Wilson lines in (\ref{tensor2})
 cancel each other and the gauge transformation reduces to
(\ref{gauge-trans}).

A few comments are in order.
For a single index representation, the gauge transformation law reduces to the 
 normal NC gauge transformation
\begin{eqnarray}
 \f^i\rightarrow (U*W_C^{-1})^j_{~l}*W_C{}^l_{~k}*\f^k=U^i_{~j}*\f^j\,.
\end{eqnarray}
since the Wilson lines cancel.

The gauge transformation for the rank-2 tensor $\phi^{ij}$
 (\ref{tensor2}) cannot be decomposed into symmetric and antisymmetric
 representations like as commutative case
 since the gauge transformation does not commute with the
 interchange of the indices
\begin{eqnarray}
(\phi^U)^{ij}\longrightarrow
 (\phi^U)^{ji}=(U*W_{C_2}^{-1})^i_{~k}*U^j_{~l}*W_{C_2}{}^k_{~m}*\phi^{lm}\not
 =(U*W_{C_2}^{-1})^j_{~k}*U^i_{~l}*W_{C_2}{}^k_{~m}*\phi^{ml}. \nonumber \\
\end{eqnarray}
In other words, in the NC case rank-2 tensor is not reducible, and we
 cannot treat $\phi^{(ij)}={1\over 2}(\phi^{ij}+\phi^{ji})$ as symmetric
 representation (similarly to antisymmetric case).
Instead of it, one can construct the following symmetric gauge
 invariant tensor
\begin{eqnarray}
 \Phi^{(ij)}=W_{C_1}{}^{(i}_{~a}*W_{C_2}{}^{j)}_{~b}*\phi^{ab}
 ={1 \over 2}(W_{C_1}{}^i_{~a}*W_{C_2}{}^j_{~b}
+W_{C_1}{}^j_{~a}*W_{C_2}{}^i_{~b})*\phi^{ab}\,,\label{sym}
\end{eqnarray}
where $\phi^{ab}$ follows the gauge transformation law (\ref{tensor2})
 while $\Phi^{(ij)}$ is a symmetric tensor.
The antisymmetric tensor $\Phi^{[ij]}$ is given by
\begin{eqnarray}
 \Phi^{[ij]}=W_{C_1}{}^{[i}_{~a}*W_{C_2}{}^{j]}_{~b}*\phi^{ab}
 ={1 \over 2}(W_{C_1}{}^i_{~a}*W_{C_2}{}^j_{~b}
 -W_{C_1}{}^j_{~a}*W_{C_2}{}^i_{~b})*\phi^{ab}\, .
\end{eqnarray}

Similarly,
one can define the modified gauge transformation for
fermions.
For example, the gauge transformation
for the fermionic 2-tensor $\psi^{ij}$ is given by
\begin{eqnarray}
 \psi^{ij}\rightarrow
 (\psi^U)^{ij}
=(U*W_{C_2}^{-1})^j_{~k}*U^i_{~l}*W_{C_2}{}^k_{~m}*\psi^{lm}\,,
\end{eqnarray}
corresponding to the gauge invariant quantity
$ \Psi^{ij}\equiv W_{C_1}{}^i_{~i^\prime}*W_{C_2}{}^j_{~j^\prime}*\psi^{i^\prime
  j^\prime}$.
Hereafter we will restrict our attention to scalar fields as it is straightforward to
apply for fermions.

\subsection{Fields charged under an arbitrary number of $U_*(N)$ groups}\label{direct-product}

Now we extend the above discussion into construction of representations
 for a direct product of any number of $U_*(N)$
 with different $N$.
We start by considering a direct product of two groups $U_*(M)\times
 U_*(N)$ and a field charged under these two factors, $\f^{mn}$ where
 $m$ and $n$ denote gauge indices for fundamental representations of
 $U_*(M)$ and $U_*(N)$, respectively.
Performing the simple NC version of
 gauge transformation for fundamental
 representation for the gauge group $U_*(M)\times U_*(N)$, we have
\begin{eqnarray}
 \f^{mn}\rightarrow (\f^{mn})^U=(U_N)^n_{~n^\prime}
 *(U_M)^m_{~m^\prime}*\f^{m^\prime n^\prime}\,. \label{trans-direct1}
\end{eqnarray}
This does not satisfy the closure condition (\ref{times}), i.e.,
$((\f^{mn})^U)^V\neq(\f^{mn})^{V*U}$.

Thus, we would like to modify the gauge transformation law 
 so that it satisfies (\ref{times}) similarly to the case of the tensorial
 representation for a single $U_*(N)$ gauge group.
As in (\ref{tensor2-inv}) of the previous subsection,
 we require gauge invariance of a quantity,
\begin{eqnarray}
 \Phi^{mn}=(W_M)^m_{~m^\prime}
 *(W_N)^n_{~n^\prime}*\phi^{m^\prime n^\prime}\,, \label{tensord-inv}
\end{eqnarray}
where $W_M$ and $W_N$ are the Wilson lines for the gauge group $U(M)$
 and $U(N)$, respectively.

Each Wilson line has a contour with end points as in Eq. (\ref{path})
 and follows the gauge transformation law (\ref{W-trans}).
The exact shape of the contour may be different for $M$ and $N$.
We then define the gauge transformation law so that (\ref{tensord-inv}) is
 gauge-invariant:
\begin{eqnarray}
 \f^{mn}\rightarrow
 (\f^U)^{mn}=(U_N*W_N^{-1})^n_{~k}*(U_M)_{~l}^{m}*(W_N)^{k}_{~p}*\f^{l
  p}\,. \label{trans-direct2}
\end{eqnarray}
For notational convenience, we write this in the tensor notation:
\begin{eqnarray}
 \f^U=(1\otimes U_N*W_N^{-1})*(U_M\otimes W_N)*\f\
\end{eqnarray}
This gauge transformation law satisfies the group multiplication law
 (\ref{times}), and it is therefore a suitable NC gauge transformation.
Note that this transformation law includes only the Wilson line
 $W_N$ not but $W_M$.

There is another possible form of the gauge invariant object
\begin{eqnarray}
 \Phi^{mn}=(W_N)^n_{~n^\prime}
 *(W_M)^m_{~m^\prime}*\phi^{m^\prime n^\prime}\,. \label{direct-inv2}
\end{eqnarray}
Gauge transformation associated with (\ref{direct-inv2}) is given by
\begin{eqnarray}
 \f^{mn}\rightarrow
  (\f^{mn})^U=(U_M*W_M^{-1})^m_{~k}*(U_N)_{~l}^{n}*(W_M)^{k}_{~p}*\f^{p
  l}\,. \label{trans-direct3}
\end{eqnarray}
This gauge transformation also satisfies the closure condition
 (\ref{times})
 and in this case includes the Wilson line
of the $U_*(M)$ group not $U_*(N)$.
Both of the gauge transformation (\ref{trans-direct2}) and
 (\ref{trans-direct3}) fall into the ordinary gauge transformation
 in the commutative limit.
Now the noncommutativity seems to split
 the ordinary-space representation into two distinct NC representations.
However, as we will see in the next section,
 they would lead to the same physical result.
Therefore, in the following, we will adopt (\ref{trans-direct2}) .

We generalize the representation (\ref{trans-direct2}) into one
 for a direct product of $n$ unitary gauge groups.
In tensor notation, we obtain the gauge transformation as
\begin{eqnarray}
  \f_{[n]}^U&=&(U_{M_n}*W_{M_n}^{-1}\otimes 1 \otimes \cdots\otimes
  1)*(1\otimes U_{M_{n-1}}*W_{M_{n-1}}^{-1}\otimes 1\otimes \cdots
  \otimes 1)
\nonumber \\
 &&*\cdots *(1\otimes \cdots \otimes U_{M_1}*W_{M_1}^{-1})
*(W_{M_1}\otimes \cdots \otimes
 W_{M_n})*\f_{[n]}\,, \label{direct-general}
\end{eqnarray}
where $W_{M_i}$ is the Wilson line for $U(M_i)$ gauge group.
If $\phi$ has any anti-fundamental indices, they can be taken to
 transform
 from the right.
Thus we have obtained the field $\phi_{\left[n\right]}$ carrying
 $n$ charges.
Here the corresponding gauge invariant object is
\begin{eqnarray}
 \Phi_{\left[n\right]}
 =W_{M_1}\otimes W_{M_2}\otimes \cdots \otimes W_{M_n}*
  \phi_{\left[n\right]}\,. \label{direct-inv}
\end{eqnarray}
For anti-fundamental indices one
 finds a similar gauge invariant quantity by multiplying
 with the corresponding Wilson lines from the right
 instead of the left.
In this case, it is also easy to see that taking the commutative limit, the gauge
 transformation (\ref{direct-general}) reduces to the commutative one
 since the Wilson lines cancel.

%
%
\sect{Gauge Invariant Action Integral}

Now that we have obtained fields carrying any number of charges,
 let us construct a gauge invariant action integral.
In what follows we  focus on the
 rank-2 representation (\ref{trans-direct2})
 for a direct product of two gauge groups.
We introduce the gauge invariant vector field,
\begin{eqnarray}
 {\cal A}_\mu^{L} \equiv A_\mu^{W_{L}}
 =W_{L}*(A_\mu^{L} - i\partial_\mu)*W_{L}^{-1}\,,
\end{eqnarray}
where $L=(M,N)$.
With this vector field, we define the covariant derivative:
\begin{eqnarray}
 {\cal D}_\mu^{MN}=(1\otimes 1)\partial_\mu+i({\cal A}_\mu^M \otimes
  1)+i(1\otimes {\cal A}_\mu^N)\,. \label{covariant derivative}
\end{eqnarray}
By using them, we can write down the gauge invariant kinetic term:
\begin{eqnarray}
 S=\int d^4 x~{\rm tr}|{\cal D}_\mu^{MN}*\Phi|^2
 =\int d^4 x~{\rm
  tr}|D_\mu^{MN}*\phi|^2\,, \label{direct-kinetic}
\end{eqnarray}
where
\begin{eqnarray}
 D_\mu^{MN}=(W_M \otimes W_N)^{-1}*((1\otimes 1)\partial_\mu +({\cal
  A}_\mu^M\otimes 1)+(1 \otimes {\cal A}_\mu^{N}))*(W_M\otimes W_N)\,.
\end{eqnarray}
This gauge invariant kinetic term is an NC extension
 of the commutative gauge invariant kinetic term for the fundamental
 representation for direct product of the two gauge groups.
Taking the $\theta^{\mu\nu}\to 0$ limit in (\ref{direct-kinetic}), the Wilson lines
 cancel and it reduces to the
 ordinary gauge invariant kinetic term
\begin{eqnarray}
 S=\int d^4 x {\rm tr}|\{(1\otimes 1)\partial_\mu
  +i(A_\mu^M\otimes 1)+i(1\otimes A_\mu^N)\}\phi|^2\,. \label{cc kinetic term}
\end{eqnarray}

Recall that the Wilson line is a gauge group element.
From the gauge transformation (\ref{trans-direct2}),
 one sees that the gauge invariant quantity (\ref{tensord-inv})
 can be expressed in terms of a
 gauge transformation with $U_M=W_M$ and $U_N=W_N$:
\begin{eqnarray}
 \Phi=\phi^{U_M=W_M,U_N=W_N}\,.\label{inva}
\end{eqnarray}
The gauge invariant field is obtained through a gauge transformation
 with $U(x)=W(x)$.
This means that $\Phi$ and ${\cal A}_\mu^{M(N)}$ lie on the same gauge
 orbit with $\f^{mn}$ and $A_\mu^{M(N)}$, respectively.
Therefore the gauge invariant kinetic term is of the same form as the
 ordinary-space one but with the Moyal star-product between any field.
This is actually a gauge fixing procedure as indicated
 for the tensorial representation for a single simple gauge group
 in Ref. \cite{Chu:2002}.
Thus as we mentioned, (\ref{direct-inv2}) is physically
 equivalent to (\ref{tensord-inv}).
It is also straightforward to obtain
 the action integral of gauge theory coupled to
 the rank-$n$ field $\phi_{\left[n\right]}$
 with the gauge transformation (\ref{direct-general})
 and to fermionic rank-$n$ fields.

Considering the fact that $\Phi^{mn}$ itself is gauge invariant, one can easily
 construct other gauge invariant candidates for the kinetic term, for example:
\begin{eqnarray}
S=\int d^4 x \partial_\mu \Phi^{mn}*\partial^\mu \Phi_{mn}^\dagger\,.
 \label{kinetic 2}
\end{eqnarray}
After the gauge fixing procedure described above, this would give
 the usual kinetic
 term of a gauge invariant NC scalar field.
On the other hand there is no obvious theoretical reason to
 prefer the (invariant) covariant derivatives (\ref{covariant derivative})
 in construction of the kinetic term as
 the composite field $\Phi$ is actually invariant and one could do simply
 with ordinary derivatives.
We simply adopt the covariant derivatives
 (\ref{covariant derivative}) for
 phenomenological applications.

Finally, we would like to make a comment concerning the
 so-called Higgsac mechanism
 \cite{CKT,AST}.
As we explained in the introduction, the minimal NC extension of the SM
 gauge group is $U_*(3)\times U_*(2)\times U_*(1)$.
In order to realize the commutative SM at low energies, one has to break the
 trace $U_*(1)$ parts of these groups.
The Higgsac mechanism was proposed to realize such a breaking
 in a manner that respects unitarity and the requirements of the no-go theorem.
The basic ingredient of this mechanism is a scalar field
 $\phi_{[n]}=\phi^{i_1i_2\cdots i_n}$ that is
 a rank-$n$ tensor under the $U_*(N)$ gauge group
 (whose extension into the case of direct product of
 any number of gauge group is straightforward,
 following the construction explained in section \ref{direct-product}).
Similar to the case of rank-$2$ tensor in section \ref{tensor-section},
 the gauge transformation for the rank-$n$
 tensor $\phi_{[n]}$ and the associated gauge invariant
 object are given as 
\begin{eqnarray}
 (\phi_{[n]})^U
 &=&(U* W^{-1}\otimes \cdots \otimes 1)*(1\otimes U* W^{-1}
 \otimes \cdots \otimes 1) \nonumber \\
 &&*\cdots *(1\otimes \cdots \otimes U* W^{-1})*\phi_{[n]}\,,
\end{eqnarray}
and
\begin{eqnarray}
 \Phi={1 \over n!}\epsilon_{i_1 i_2...i_n}
  W^{[i_1}_{~j_1}*W^{i_2}_{~j_2}*\cdots
  *W^{i_n]}_{~j_n}*\phi^{j_1j_2\cdots j_n}\,. \label{higgsac1}
\end{eqnarray}
The latter is called the Higgsac field.
Here we have used the same Wilson line for simplicity differently from
 the discussion in section 
\ref{tensor-section}.
With the use of the field (\ref{higgsac1}) it was suggested that
 the following Lagrangian caused
 a spontaneous breaking of the trace $U_*(1)$ part of the $U_*(N)$
\begin{eqnarray}
{\cal L}&=&\partial_\mu \Phi^\dagger * \partial^\mu\Phi
 + m^2 |\Phi|^2 -{\lambda \over 2} |\Phi|^4\,,
\label{higgsacaction}
\end{eqnarray}
In this Lagrangian, the scalar field $\Phi$ has
 a non-zero vacuum expectation value
 $\langle\Phi \rangle=\langle \phi \rangle=\sqrt{m^2/\lambda}$
 where $\phi\equiv {1 \over n!}\epsilon_{i_1i_2\cdots
 i_N}\phi^{[i_1i_2\cdots i_n]}(x)$ and $A_\mu=0$.
Expanding $\Phi$ with respect to $\theta$ and the gauge coupling
 constant, one finds
\begin{eqnarray}
 \partial_\mu \Phi=
  (\partial_\mu + i n g  A_\mu^0)\phi
 + ig \partial_\mu \phi \int_0^1 d\sigma {d\zeta^\mu \over d\sigma}
 A_\mu^0(x+\z(\sigma))
 + {\cal O}(\theta)+{\cal O}(g)\,,
\end{eqnarray}
where $A^0_\mu$ is the trace part of $A_\mu$.
From this expression, it appears that the gauge field $A_\mu^0$ has a
 mass in the presence of the non-zero vacuum expectation value of $\Phi$.

However, according to our discussion above,
 all the gauge fields included in the scalar field $\Phi$
 are gauged away by fixing the gauge.
Therefore, no coupling between the scalar field
 and the gauge field occurs
 and there cannot exist any mass term for symmetry breaking.
Thus the symmetry breaking proposed in \cite{CKT, AST}
 would be an artifact of using a truncated expansion.

\sect{Conclusion}
We have proposed a possible way out of the
 restrictions in the no-go theorem of NC gauge theories.
We have constructed fields carrying charges of
 any number of $U_*(N)$ factors.
A key ingredient for achieving such a representation is to
 satisfy the closure condition
 by modifying the gauge transformation using the half-infinite NC Wilson line \cite{Chu:2002}.
We have constructed the action integral formed out of fields
 carrying any number of charges.
The resultant action is of the same form as the
 ordinary-space one but with the Moyal star-product between any field,
 taking into account that the Wilson lines in the gauge invariant
 quantity are gauged away.
This fact leads to a result that
 within this construction
 the Higgsac mechanism discussed in \cite{CKT,AST} would not work.
One of key issues in the Higgsac mechanism is that there are
 interactions between gauge fields in the
 Wilson line and scalar fields and trace $U_*(1)$ part of the gauge
 field acquire a mass upon the condensation of the scalar fields.
However, the Wilson lines are completely gauged away.
The gauge fields no longer couple to the scalar giving a mass
 and no gauge symmetry breaking occurs.

It is interesting that, although the no-go theorem for
 noncommutative gauge fields can be circumvented to a great extent,
 there is one aspect - the construction of representations of the
 $U_*(1)$ subgroup of $U_*(N)$ gauge group - which cannot be solved.
We believe that this is connected to the fact that the quotient
 $U_*(N)/U_*(1)$ does not exist. If one could construct a
 representation of the subgroup $U_*(1)$, then by a correspondingly
 charged field one could break spontaneously the $U_*(1)$ subgroup.
However, after such a breaking, there is no noncommutative gauge
 symmetry left, since $SU_*(N)$ does not exist. These aspects
 strongly remind the situation encountered in quantum groups, when
 upon the deformation, a subalgebra of an algebra is no more
 subalgebra of the deformed algebra.

\vspace*{10mm}
\begin{center}
{\bf Acknowledgements}
\end{center}
We are grateful to Masud Chaichian and Peter Pre\v{s}najder for
 useful discussions. The work of N.U. is supported by Bilateral
 exchange program between Japan Society for the Promotion of Science
 and the Academy of Finland. S.S. acknowledges a grant from GRASPANP,
 the Finnish Graduate School in Particle and Nuclear Physics.

%
%
\small{

\end{document}